\documentstyle[prl,aps,epsfig]{revtex}
\begin{document}
\draft
\twocolumn[\hsize\textwidth\columnwidth\hsize\csname@twocolumnfalse%
\endcsname
\preprint{\parbox[t]{45mm}{\small  UNITU-THEP-08/01\\ hep-ph/0102332 \\}}

\title{Verifying the Kugo--Ojima Confinement Criterion in Landau Gauge 
Yang--Mills theory}

\author{Peter Watson and Reinhard Alkofer}

\address{Institute for Theoretical Physics, T\"ubingen University, 
Auf der Morgenstelle 14, D-72076 T\"ubingen, Germany}
\date{\today}
\maketitle
\begin{abstract}
Expanding the Landau gauge gluon and ghost two--point functions in a power
series we investigate their infrared behavior. The corresponding powers are
constrained through the ghost Dyson--Schwinger equation by exploiting 
multiplicative renormalizability. Without recourse to any specific truncation
we demonstrate that the infrared powers of the gluon and ghost propagators are
uniquely related to each other. Constraints for these powers are derived, and 
the resulting infrared enhancement of the ghost propagator signals that the 
Kugo--Ojima confinement criterion is fulfilled in Landau gauge Yang--Mills 
theory.
\end{abstract}

]

Modifications to the standard framework of quantum field theory are necessary
to accommodate the confinement phenomenon in QCD. Here either relaxing the
principle of locality or abandoning the positivity of the representation space
(or both) is required. As quantum gauge field theories in covariant gauges
demand indefinite metric spaces, and as locality is of fundamental importance,
it is suggestive to relate confinement to the violation of positivity in 
Yang--Mills theories:
No colored states should be present in the positive definite space of physical
states defined by some suitable condition maintaining physical $S$-matrix
unitarity. Exploiting the BRS symmetry of the gauge fixed action a nilpotent
BRS-charge $\mathcal{Q_B}$ in an indefinite metric space $\mathcal{V}$ can be
constructed. The subspace $\mathcal{V_{\mbox{\footnotesize p}}} = \mbox{Ker}\,
Q_B$ is defined by those states which are annihilated by the BRS charge
$\mathcal{Q_B}$. Positivity is then proved for physical states
\cite{Kugo:1979gm} (see also refs.\ \cite{Nakanishi:1990qm,Alkofer:2000wg})
which are given by the cohomology $\displaystyle \mathcal{H}(Q_B,\mathcal{V}) =
{\mbox{Ker}\, Q_B}/{\mbox{Im} Q_B} $, the covariant space of equivalence
classes of BRS-closed modulo BRS-exact states.

Longitudinal and timelike gluons form BRS quartets together with ghosts and are
thus unphysical. At the same time the global symmetry $J_{\mu ,\nu}^a$
corresponding to gauge transformations generated by $\theta^a(x) = a^a_\mu
x^\mu$ is spontaneously broken quite analogous to the displacement symmetry in
QED. The identification of BRS-singlet states with color singlets is 
non-perturbatively  possible if this global symmetry is restored 
\cite{Kugo:1979gm,Nakanishi:1990qm,Kugo:1995km}. The condition for a dynamical
restoration of this symmetry is the Kugo--Ojima confinement criterion which 
is a requirement on the infrared behavior of a four-point Green's function. 
In Landau gauge, a sufficient condition for this criterion is that
the nonperturbative ghost propagator is more singular than a massless pole in
the infrared \cite{Kugo:1995km}: 
\begin{equation} 
D^{ab}_{G}(p) = \delta^{ab} \frac {G(p^2)}{p^2} \, , \;\; \mbox{with} 
\quad G(p^2) \stackrel {p^2 \to 0 } {\longrightarrow}\infty \, . 
\label{GhProp} 
\end{equation} 
This mechanism is correlated to the derivation~\cite{Nishijima:1996ji} of the
Oehme--Zimmermann superconvergence relations~\cite{Oehme:1980ai} from
Ward--Takahashi identities. These superconvergence relations formalize a long
known contradiction between asymptotic freedom and the positivity of the
spectral density for transverse gluons in the covariant gauge.

The violation of positivity of transverse gluons is unambiguously established
by a variety of independent nonperturbative studies of the gluon propagator
\cite{Mandula:1999nj}. Furthermore, recent lattice calculations 
investigate  the Kugo--Ojima criterion directly \cite{Furui:2000mq}. As
infrared singularities are anticipated, also non-perturbative {\em continuum}
methods, besides lattice calculations, are necessary to investigate this and
related pictures of confinement in more detail.

The Green's functions of a quantum field theory obey the Dyson--Schwinger
equations. For realistic theories this infinite tower of integral equations
cannot be solved directly, and truncations are necessary to obtain (numerical)
solutions. The ones for QCD propagators can then be used to formulate a
successful hadron physics phenomenology, see {\it e.g.} the recent reviews
\cite{Alkofer:2000wg,Roberts:2000aa}.

The results of two recent truncation schemes
\cite{vonSmekal:1997is,Atkinson:1998tu} suggest that the Kugo--Ojima
confinement criterion for the ghost propagator and the violation of positivity
for transverse gluon correlations are fulfilled. Obviously, the question arises
whether these results are sensitive to the employed truncation schemes. 

In this letter we will exploit the general structure of the simplest QCD
Dyson--Schwinger equation, the one for the ghost two--point function, to
relate  the infrared behavior of gluon and ghost propagators. Thereby {\em the 
Kugo--Ojima confinement criterion will be verified by purely analytical
methods}. Using the rather unique properties of the ghost-gluon vertex in the
Landau gauge and assuming the applicability of a powerlaw ansatz in the
infrared the ghost Dyson--Schwinger equation admits a
completely general qualitative analysis. Instead of truncating the system, the
general forms for the unknown functions constrained by various consistency
arguments, notably multiplicative renormalizability, are input into the
equation under an asymptotic expansion in the infrared momentum scale, {\it
c.f.\/} the method described in ref.\ \cite{Pagels:1977xv}.

We start by providing some basic notations. 
The general form of the gluon propagator in Landau gauge is purely transverse
and is given by \cite{Alkofer:2000wg,Roberts:2000aa}
\begin{equation}
D^{ab}_{\mu\nu}(p) = \delta^{ab} \left( g_{\mu\nu} - \frac{p_\mu p_\nu }{p^2}
\right) \frac {Z(p^2)}{p^2} \; ,
\label{GlProp}
\end{equation}
which defines the gluon renormalization function $Z(p^2)$.
The ghost renormalization function $G(p^2)$ is established in eq.\
(\ref{GhProp}). We define the ghost-gluon vertex 
\begin{equation}
\Gamma_\mu^{abc} (p_1,p_2,p_3) = - ig f^{abc}
\Gamma_\mu (p_1,p_2,p_3)
\end{equation}
with antighost ($p_1$), ghost ($p_2$) and gluon ($p_3$) momentum incoming, {\it
i.e.} $p_1+p_2+p_3=0$. At tree-level the dressing functions reduce to
\begin{equation}
Z(p^2)=G(p^2)=1, \quad \Gamma_\mu (p_1,p_2,p_3) = p_{1\mu} \; .
\end{equation}
These renormalized dressing functions are related to the bare ones by the usual
renormalization constants $Z_3$ (gluon wave function), $\widetilde Z_3$ 
(ghost wave function)  and $\widetilde Z_1$ (ghost-gluon vertex). 
In Landau gauge one has $\widetilde Z_1 =1$ 
\cite{Taylor:1971ff,Marciano:1978su}, {\it i.e.}, the ghost-gluon vertex
has not to be renormalized .

The renormalized ghost dressing function $G(p^2)$ is a dimensionless function
of a single variable. In pure Yang--Mills theory there are only two scales
present: the external scale $p^2$ and the renormalization scale $\mu^2$.
Therefore $G$ is a function of the ratio $p^2/\mu^2$ only. As the external
scale vanishes this function can be written as an asymptotic expansion.
Renormalizability implies that it must also depend on the renormalized coupling
$g$. This motivates the power series ansatz
\begin{equation}
G(p^2;\mu^2) = \sum _{n} d_n \left( \frac {p^2}{\mu^2} \right) 
^{\delta_n} \; ,
\label{Gseries}
\end{equation}
where the coefficients $d_n$ and the exponents $\delta_n$ are, at least in
principle, functions of $g$. We will see below that the exponents
$\delta_n$ will be independent of $g$. Furthermore, as will be shown later on,
the series (\ref{Gseries}) has a lowest power $\delta_0$ such that as the
external scale $p^2$ vanishes the ghost dressing function $G(p^2)$ is well
approximated by a powerlaw. Multiplicative renormalizability allows to derive
the renormalization group (RG) equation
\begin{equation} 
\left( \mu \frac{\partial }{\partial \mu } + \beta (g)
\frac{\partial }{\partial  g} -2 \gamma_{\rm G} \right)  G(p^2;\mu^2) = 0
\label{CS}
\end{equation} 
with $\beta (g)$ being the beta function, and 
\begin{equation}
\gamma_{\rm G} = \mu \frac{\partial \log \widetilde Z_3 ^{-\frac 1 2}}
{\partial \mu } 
\end{equation} 
is related to the ghost anomalous dimension. Inserting the series 
(\ref{Gseries}) into eq.\ (\ref{CS}), and exploiting that $p^2$ is a free 
(though small) parameter one obtains 
\begin{eqnarray} 
-d_n\delta_n + \beta (g) \left( \frac{\partial  d_n }{\partial  g}  + 
d_n \log  ( \frac {p^2}{\mu^2} )
\frac{\partial \delta_n }{\partial  g} \right) +
d_n \gamma_{\rm G}   = 0.  
\end{eqnarray} 
Since the logarithm has no matching factor the exponents $\delta_n$ must be 
independent of the renormalized coupling 
(${\partial \delta_n }/{\partial  g} = 0$).
Using $\beta (g) = - g (2\gamma_{\rm G} + \gamma_{\rm A} )$ with 
$\gamma_{\rm A} = \mu {\partial \log Z_3^{-\frac 1 2}}/{\partial \mu }$ 
the resulting differential equation for the coefficients $d_n$ has the general 
solution 
\begin{equation} 
d_n = {\rm const.} \; g^{-\frac{2(\delta_n + \gamma_{\rm G})}{
2\gamma_{\rm G} + \gamma_{\rm A}}} . \label{dnsol} 
\end{equation}
Expanding the renormalized gluon dressing function 
\begin{equation}
Z(p^2;\mu^2) = \sum _{n} e_n \left(\frac {p^2}{\mu^2}  
\right) ^{\epsilon_n}
\label{Zseries}
\end{equation}
one arrives in a completely analogous way at
\begin{equation}
e_n = {\rm const.} \; g^{-\frac{2(\epsilon_n + \gamma_{\rm G})}
{ 2\gamma_{\rm G} + \gamma_{\rm A}}} .
\end{equation}

In the renormalization process we have to specify the corresponding
normalization conditions. We require that the unrenormalized functions reduce
to a constant at asymptotically large momenta. Exploiting additionally
multiplicative renormalizability and using eq.\ (\ref{Gseries}) one obtains
the formal relation
\begin{equation}
\widetilde Z_3(\mu^2,\Lambda ^2) \propto \left( 
 \sum _{n} d_n \left( \frac {\Lambda ^2}{\mu^2} \right) ^{\delta_n}
 \right) ^ {-1} .
\label{tZ3series}
\end{equation}
This equation is only well-defined if the corresponding series makes sense, {\it
e.g.\/} if it is resumable by some method. This is the main assumption
throughout the presented derivation. One consistency check is provided by
inserting the general solution for the coefficients $d_n$ (\ref{dnsol})
and rewriting the renormalized coupling $g$ in terms of the scale $\mu$.
This yields
\begin{equation}
\widetilde Z_3(\mu^2,\Lambda ^2) \propto \left( \frac {\Lambda ^2}{\mu^2}
\right) ^{\gamma_{\rm G}}
\label{Z3p}
\end{equation}
which is precisely the correct dependence to agree with the definition of
$\gamma_{\rm G}$. A completely analogous argument applies to the gluon
renormalization constant $Z_3$.

The tensor structure  of the ghost-gluon vertex reads
\begin{equation}
\Gamma_{\mu}(p_{1},p_{2},p_{3})=p_{1\mu}A(p_{i}^{2})+
                                p_{3\mu}B(p_{i}^{2}) \; .
\label{ghgl1}
\end{equation}
In Landau gauge the ghost-gluon vertex becomes bare when the
ghost momentum $p_2$ vanishes \cite{Taylor:1971ff,Marciano:1978su}:
\begin{equation}
\Gamma_{\mu}(p_{1},p_{2},p_{3})=p_{1\mu} \left( A 
- B \right) - p_{2\mu} B
  \stackrel{p_{2}\rightarrow 0}{ \longrightarrow } p_{1\mu} \; .
\end{equation} 
Thus, as $p_2\to 0$ the functions $A$ and $B$ cannot be singular. This is in
accordance with the fact that the renormalization constant $\widetilde  Z_1 =1$.

The RG equation for the vertex function can be split into two
parts, one for each of the tensor components. It is only necessary to consider
the equation for $A$, and since $\widetilde  Z_1 =1$, it is identical to the
RG equation for the coupling. Writing the vertex function as a
series has the complication of several momentum scales. For our considerations
it is sufficient to simply extract the scale and coupling dependence
\begin{equation}
A(p_{i}^{2};\mu^2)= \sum_k a_k(p_{1}^{2},p_{2}^{2}) 
\lambda_k
\left(\frac{p_{3}^{2}}{\mu ^2}\right)^{\kappa_{k}} ,
\label{AR}
\end{equation} 
with
\begin{equation}
\lambda_k =  
\left( g^2 \right)^{-\frac{\kappa_{k}}{\gamma_{\rm A}+2\gamma_{\rm G}}} \; .
\end{equation} 
We must renormalize the ghost-gluon vertex at the point $p_{2}=0$ since this
corresponds to a zero finite renormalization from $\widetilde Z_1 =1$. 
Knowing the $p_{2}\rightarrow0$ limit of the full vertex
does not constrain $A$. (When $p_{2}=0$ one only has $A-B=1$.)
With this given limit and the additional requirement that the ghost equation
should provide a consistent solution one infers that
\begin{eqnarray}
A(p_1^2=p_3^2=\Lambda^2,p_2^2=0;\mu^2) =
 \sum_k a_k \lambda_k
\left(\frac{\Lambda^2} {\mu^2} \right)^{\kappa_{k}} 
\label{Ap} 
\end{eqnarray} 
is a finite positive constant.
Asymptotic freedom requires that $\gamma_{\rm A}+2\gamma_{\rm G} > 0$. In the 
unrenormalized theory the vertex should be bare when the unrenormalized
coupling vanishes because $\widetilde Z_1 =1$. Therefore $\kappa_{k}\leq0$ with
the highest value  $\kappa_{0}=0$. 

The ultraviolet region of the ghost Dyson--Schwinger equation corresponds to a
small (albeit non-vanishing) ghost momentum $p_2$ and all other momenta large.
For the following discussions it is helpful to write
\begin{eqnarray}
A(p_{i}^{2};\mu^2) _{p_2\to 0} = 
\sum_{k,m} b_k(p_1^2) \lambda_k \left(\frac{p_3^2}
{\mu^2} \right)^{\kappa_{k}} \left(\frac{p_{2}^{2}}{p_{3}^{2}}\right)^{\rho_{m}}
\label{Aseries}
\end{eqnarray}
where the $\rho_{m}\geq0$ and $\rho_{0}=0$.

The full Euclidean ghost Dyson--Schwinger equation can be written after
having performed the two trivial angular integrals as
\begin{eqnarray}
G^{-1}(x) = \widetilde Z_3 && \\
- \frac {C_A g^2}{16\pi^2} \int _{0}^{\Lambda^2}
dy Z(y) &&\frac 2 \pi \int_0^\pi d\theta \sin^4\theta 
\frac {G(z)}{z} A(z,x,y) \; , \nonumber
\end{eqnarray} 
where $\theta$ is the angle between the two internal loop momenta corresponding 
to $y=q^2$ and $z=(p-q)^2$ with $x=p^2$. $C_A=N_c$ is the resulting color
factor.
When using the expansions (\ref{Gseries}) and (\ref{Zseries})   the angular
integration splits the radial integral into two parts,  the lower part $I_{gh}$ 
($0\rightarrow x$) and the upper part $J_{gh}$ ($x\rightarrow \Lambda^2$). The
expression derived from  the upper limit will be some combination of factors
with the ratio $x/\Lambda^2$ to some power.  The effect of an exact angular
integration is to generate a hypergeometric function whose argument is $x/y$
and which can be expanded in a convergent series with positive integer powers
of $x/y$ starting with unity.  Thus, the $z$-dependence of the vertex function
is unimportant.

The most important aspect is that the ghost-gluon vertex reduces to its
tree-level form as the in-ghost momentum vanishes. In the ultraviolet part of
the integral this is precisely the case: the in-ghost momentum is the infrared
external scale $x$.  Therefore the vertex function can be written in the form
of eq.\ (\ref{Aseries}).

Using the series for $G$ and $Z$, it is possible to write the ghost equation 
with the most general vertex function as
\begin{eqnarray}
\left( \sum _{n} d_n \left( \frac {x}{\mu^2} \right) ^{\delta_n}
\right) ^ {-1} = \widetilde Z_3 &&\label{Gh}\\
 - \frac {C_A g^2}{16\pi^2} 
\sum _{j,k} e_j d_k \left( \frac {x}{\mu^2} \right) ^{\epsilon _j + \delta _k}
&&\Big(I_{gh}(\epsilon _j,\delta _k) + J_{gh}(\epsilon _j,\delta _k)\Big)
\; . \nonumber 
\end{eqnarray}
where $I_{gh}$ and $J_{gh}$ represent the lower and upper regions of the
integral, respectively. 
The integrals $I_{gh}$  must contain factors characterized by the $\kappa_k$
but there are no restrictions on its form except that the vertex must have 
positive powers of the coupling. As the integral $I_{gh}$ is very difficult to
compute we will not be able to proceed by studying its general properties.

It is frequently asserted that only the infrared integrals of  Dyson--Schwinger
equations can contribute to the renormalized propagator  functions on the left
hand side because this is the only fully renormalizable possibility. Indeed,
one can show that upper integral $J_{gh}$ will not contribute directly to the
propagator functions \cite{Watson:2001}. It is the renormalization constant
$\widetilde Z_3 $ that provides the key to the infrared behavior of the
propagators, and only the ultraviolet integral can yield the ultraviolet
divergent part of the ghost equation.  First we note that $J_{gh}(\epsilon
_j,\delta _k)$ can be expanded as
\begin{eqnarray}
J_{gh}=
\sum_{l,m,q} J_{j,k,l,m,q} 
\left( \frac {\Lambda^2}{x} \right) ^{\epsilon _j + \delta _k - \rho_m - q}
\lambda_l\left(\frac{\Lambda^2}
{\mu^2} \right)^{\kappa_{l}}&& \; ,
\label{Jgh}
\end{eqnarray}
where $\rho_{m}\geq0$ represents the kinematical content of the vertex function,
and $q=0,1,\ldots$ denotes the order of the angular integration series.
As $\widetilde Z_3$ has to be independent of $x$ one deduces
from eqs.\ (\ref{Gh},\ref{Jgh}) that
only $\rho_0=q=0$ will contribute. All other terms
involve $x/\Lambda^2$ to a positive power and subsequently vanish:
\begin{eqnarray}
\widetilde Z_3 (\mu^2,\Lambda^2) &=& \nonumber \\
 \frac {C_A g^2}{16\pi^2} &&
\sum_{j,k,l} e_jd_kJ_{j,k,l,0,0} \lambda_l 
\left(\frac{\Lambda^2}{\mu^2} \right)^{\epsilon _j + \delta _k+\kappa_{l}} .
\end{eqnarray}
Using that expression (\ref{Ap}) is a constant, 
employing the expansion (\ref{tZ3series}) and multiplying through 
with the left hand side yields
\begin{equation}
\frac {C_A g^2}{16\pi^2}
\sum_{i,j,k} e_jd_id_kJ_{0,0,0}(\epsilon _j,\delta _k)
\left(\frac{\Lambda^2}{\mu^2} \right)^{\epsilon _j + \delta _i + \delta _k} 
= {\rm const.}
\label{one}
\end{equation}  
Since $\epsilon _0$ and $\delta _0$ are the unique lowest powers, and the
term with $i=j=k=0$ has nothing to compete with it must be responsible for the
non-vanishing part of the left hand side of this equation. Therefore it must be
independent of $\Lambda^2/\mu^2$, and we obtain
\begin{equation}
\epsilon _0 =  -2 \delta _0 \; ,
\label{central}
\end{equation}
{\it i.e.} we have shown that {\em the leading infrared powers of the gluon 
and the ghost propagator are uniquely related.}
Note that this result is valid even if corresponding series are not convergent.
We have simply
matched the coefficients of an asymptotic series. Therefore, our underlying
assumption is that the employed series have to exist in the sense of being, at
least, asymptotic ones. 

Eq.\ (\ref{one}) constrains the
values of the related leading powers. To do this, one needs to explicitly
evaluate the integrals $J_{j,k,0,0,0}$ which are readily calculated from 
the full ghost equation with a bare vertex stripping off the terms that depend
on the external scale $x$. Concentrating on the leading powers we obtain:
\begin{equation} 
- \frac 3 4 \frac {C_A g^2}{16\pi^2} e_0 d^2_0 \frac 1 {\delta _0} > 0. 
\end{equation} 
To yield a positive ghost renormalization function ({\it i.e.} to
keep the negative-norm property of the ghost states) the corresponding
leading order coefficients  $e_0$ and $d_0$ must be positive. As the 
coupling is positive this yields:
\begin{equation} 
\delta _0 < 0 \; .
\end{equation}
This relation implies that {\em the Kugo--Ojima confinement criterion}, see 
eq.~(\ref{GhProp}), {\em is fulfilled.}

A lower bound on $\delta _0$ can be derived from the lower bound of the
integral in the ghost equation. As this bound is considerably weaker  as the 
one following from a generalized
masslessness condition (see eq.\ (192) in ref.\ \cite{Alkofer:2000wg}) 
$ \delta _0>-1$,
we will not present its derivation here. This implies 
$2>\epsilon _0 >0$: The gluon propagator is suppressed in the infrared
as compared to its tree-level form.

In summary, we have investigated the general structure of the ghost
Dyson--Schwinger equation in Landau gauge. A general series form for the 
propagator and vertex functions has been employed. No specialized ansatz for
the ghost-gluon vertex has been used. Care has been taken not to use
approximations that destroy the generality of the results. Multiplicative
renormalizability, and especially the non-renormalization of the ghost-gluon
vertex, have placed the constraints to relate uniquely the leading
infrared behavior of the gluon and the ghost propagator. 

We have found an infrared enhanced ghost propagator thereby  verifying the
Kugo--Ojima confinement criterion \cite{Kugo:1979gm,Kugo:1995km} directly. Due
to the relation (\ref{central}) this corresponds to an infrared suppressed
gluon propagator. For a non-integer leading infrared  power this implies that
positivity is violated and the Oehme--Zimmermann superconvergence relation
\cite{Oehme:1980ai} is fulfilled non-perturbatively. 
Considering only the ghost equation as in this letter 
an infrared constant gluon propagator cannot be excluded. To investigate this
question one has to consider additionally the gluon Dyson--Schwinger equation
\cite{Watson:2001}. 

\smallskip

\noindent
{\bf Acknowledgements}\\
We want to express our gratitude to M.R.~Pennington who has accompanied the
whole research presented here with helpful discussions. We thank
C.\ Fischer, S.\ M.\ Schmidt and L.\ von Smekal for their valuable remarks.

\noindent
P.\ W.\ acknowledges the University of Durham and 
the CSSM, University of Adelaide,  
where major parts of this research have been done. 

\noindent
This research has been supported by DFG under contract Al 279/3-3.

\end{document}